\documentclass[aps,pra,floatfix,showpacs,noshowkeys,epsfig,graphics]{revtex4-1}%

\usepackage{graphicx}
\usepackage{amsmath}
\usepackage{amsfonts}
\usepackage{amssymb}

\newcommand{\newc}{\newcommand}
\newc{\beq}    {\begin{equation}}
\newc{\eeq}    {\end{equation}}

\newc{\beqa}    {\begin{eqnarray}}
\newc{\eeqa}    {\end{eqnarray}}
\newc{\bs}    {\section}
\newc{\no}    {\\ \nonumber}

\newc{\st}    {\stackrel}

\begin{document}
\title{ The M-sigma Relation of Super Massive Black Holes from  the  Scalar Field Dark Matter}
\author{Jae-Weon Lee}\email{scikid@jwu.ac.kr}
\affiliation{Department of Physics, North Carolina State University, Raleigh, NC 27695, USA}
\affiliation{Department of renewable energy, Jungwon university,
            85 Munmu-ro, Goesan-eup, Goesan-gun, Chungcheongbuk-do,
              367-805, Korea}

\author{Jungjai Lee}
\email{jjlee@daejin.ac.kr}
\affiliation{Division of Mathematics and Physics, Daejin University, Pocheon, Gyeonggi 487-711, Korea}

\author{Hyeong-Chan Kim}
\email{hyeongchan@gmail.com}
\affiliation{School of Liberal Arts and Sciences, Korea National University of Transportation, Chungju 380-702, Korea}

\date{\today}
\begin{abstract}
We explain the M-sigma relation between the mass of super massive black holes in
galaxies and  the velocity dispersions of their bulges  in the scalar field
or the Bose-Einstein condensate
 dark matter model.
 The gravity  of the central black holes changes  boundary conditions of the scalar field
 at the galactic  centers. Owing to the wave nature of the dark matter
  this significantly changes the galactic halo profiles even though the black holes are
  much lighter than the bulges.
As a result the heavier the black holes are, the more compact the bulges are,
 and hence the larger the velocity dispersions are. This tendency  is verified by a numerical study.
  The M-sigma relation is well reproduced with the dark matter particle mass $m\simeq 5\times 10^{-22} eV$.
\end{abstract}

 \maketitle

\section{introduction}
The M-sigma relation is a tight correlation between the
 mass $M_{bh}$ of a central super massive black hole (SMBH)  of a galaxy and
 the stellar velocity dispersion $\sigma$ in its bulge \cite{Merritt:1999ry,Ferrarese:2000se}, that is, $M_{bh}\propto \sigma^\beta$
with $\beta\simeq 4-5$.
The relation remains a mystery, because SMBHs are usually quite small and light compared to the host bulges,
and the relation is tighter than the
relation between $\sigma$ and  the mass or the luminosity of the bulge.
Furthermore, it is hard to understand how this relation survives galaxy mergers,
even if the relation was established in the early universe.
A possible solution to this problem could be a feedback mechanism acting during the galaxy
evolution. For example, Silk and  Rees \cite{Silk:1997xw} suggested that  the SMBHs
drive a wind against the accretion flow which disturbs the bulge growth, while
King \cite{King:2003ix} proposed a momentum-driven stellar wind to explain the  normalization coefficient of the relation.

However, galaxies are basically dark matter (DM) dominated objects and visible matter resides in a gravitational potential well
 generated by the DM distribution.
Considering the complicated galaxy evolution history and the varieties of galaxy types,
 it is plausible to think that the M-sigma relation is from some dynamical equilibrium conditions of the
  DM distribution not from   the properties of visible matter.

In this paper, we propose a new physical mechanism behind the M-sigma relation based on
the Bose-Einstein
condensate (BEC) or the scalar field dark matter (SFDM) model which is
proposed and studied by many authors \cite{1983PhLB..122..221B,1989PhRvA..39.4207M,sin1,myhalo,Schunck:1998nq,Matos:1998vk,PhysRevLett.84.3037,PhysRevD.64.123528,
repulsive,Fuzzy,corePeebles,Nontopological,PhysRevD.62.103517,Alcubierre:2001ea,
Fuchs:2004xe,Matos:2001ps,0264-9381-18-17-101,PhysRevD.63.125016,Julien,moffat-2006,2011PhRvD..84d3531C}.
In this model the galactic halo DM is in a BEC state of  ultra-light scalar particles with mass $m\sim 10^{-22} eV$. (For a review see \cite{2009JKPS...54.2622L,2014ASSP...38..107S,2014MPLA...2930002R,2014PhRvD..89h4040H,2011PhRvD..84d3531C,2014IJMPA..2950074H}).
This DM has a  wave-like nature \cite{2010arXiv1004.4016B} and
its large Compton wavelength suppresses small scale structure formation \cite{PhysRevD.63.063506}.
Beyond the galactic scale  the SFDM behaves like cold dark matter (CDM), and hence solves the problems of the CDM such as the missing satellites problem and
the cusp problem ~\cite{Salucci:2002nc,navarro-1996-462,deblok-2002,crisis}.
It has been also shown that the BEC/SFDM can explain the rotation curves
~\cite{PhysRevD.64.123528,0264-9381-17-1-102,Mbelek:2004ff,PhysRevD.69.127502},
  the large scale structures of the universe~\cite{2014NatPh..10..496S},
 and the spiral arms~\cite{2010arXiv1004.4016B}.

 Sin \cite{sin1} suggested that
the DM in galactic halos can be described by a
nonlinear Schr\"{o}dinger equation,
 and that the uncertainty principle
prevents halos from
 collapses.
Lee and Koh ~\cite{myhalo} proposed that
the  DM halos are giant boson stars ~\cite{review} made of
 a  complex scalar field $\bar{\psi}$ with a typical action
\beq
\label{action}
 S=\int d^4x \sqrt{-g} [\frac{c^4 }{16\pi G}R
-\frac{g^{\mu\nu}} {2} \bar{\psi}^*_{;\mu}\bar{\psi}_{;\nu}
 -U(\bar{\psi})],
\eeq where $c$ is the light velocity, and $U(\bar{\psi})=\frac{m^2 c^2}{2\hbar^2 }|\bar{\psi}|^2$ is a field potential.
 (In this paper, we  only  consider a free scalar field.)
From the action one can obtain the well-known Einstein-Klein-Gordon (EKG) equation
\beqa
\square \bar{\psi}^*&=&\frac{dU}{d\bar{\psi}}, \no
G_{\mu\nu} &=& \frac{8\pi G}{c^4} T_{\mu\nu},
\eeqa
which can describe the galactic halos made of BEC/SFDM.

The aim of this paper is to find the relation between $\sigma$ and $M_{bh}$ in the context of the BEC/SFDM.
In Sec. 2 we study the approximate profile of the BEC/SFDM halos with central BHs.
In Sec. 3  we present  results of our numerical study.
The section 4 contains discussion.

\section{Scalar field dark matter with black holes}

In the weak field limit the EKG  with a spherical symmetric metric
$
ds^2=-\left( 1+\frac{2\bar{V}}{c^2}\right) ( c d\bar{t})^2 + \left( 1-\frac{2\bar{V}}{c^2}\right) d\bar{r}^2$
is reduced to the following  Schr\"{o}dinger-Poisson equations (SPE),
\beqa
i\hbar \partial_{\bar{t}} \bar{\psi} &=&-\frac{\hbar^2}{2m} \nabla^2 \bar{\psi} +m \bar{V} \bar{\psi}, \no
\nabla^2 \bar{V} &=&\frac{4\pi G}{c^2} T_{00}=\frac{4\pi G}{c^2}( \rho_d +\rho_{vis}).
\eeqa
In this paper
 the gravitational potential $\bar{V}=\bar{V}_d-\bar{r}_s/\bar{r}$ is the sum of the  DM contribution $\bar{V}_d$ and
the black hole (BH) contribution $-\bar{r}_s/\bar{r}$.
The half of the Schwarzschild radius is $\bar{r}_s=GM_{bh}/c^2\simeq 4.78\times 10^{-14}~\frac{M_{bh}}{M_\odot}~pc$,
$\rho_d=m|\bar{\psi}|^2$ is the DM density, and $\rho_{vis}$ is the visible matter density.
The SPE can be also derived from  the mean field approximation of the BEC Hamiltonian.

  We consider a ground state (a boson star) of the SFDM with a central BH as a model of  a galaxy with mass $M$.
  This ground state model is adequate  for  small or medium size galaxies.
 The high precision numerical study~\cite{Schive:2014hza} indicates that  large galaxies in the SFDM model  consist of
 soliton-like cores and outer  tails similar to CDM profiles, thus the inner parts of the large galaxies
 resemble the ground state. For the spherical symmetric case we can ignore the outer part
 to calculate
 $\sigma$.
For simplicity we also ignore $\rho_{vis}$ in the SPE from now on.

For a numerical study it is useful to introduce  dimensionless   variables using the relations
 $\bar{r}= \hbar {r}/mc$, $\bar{t}=\hbar{t}/mc^2$, and
\beq
 \bar{\psi}=\frac{c^2}{\hbar}\sqrt{\frac{m}{4\pi G}} e^{-i\bar{E}\bar{t}/\hbar}{\psi},
 \eeq
 where $\psi$ is a real field.
Then, a dimensionless form of the SPE
 is
 \beq
 \label{SPE2}
 \left\{ \begin{array}{l}
  \nabla_{{r}} ^2 {\psi}  = 2{({V}_d-\frac{{r}_s}{{r}}-{E})}{\psi}  \\
  \nabla _{{r}} ^2 ({{V}_d -\frac{{r}_s}{{r}}}) = {\psi}^2   \\
\end{array} \right.   \\
 \eeq
where $\nabla_{{r}}^2=\partial^2_{{r}} + 2 \partial_{{r}}/{r}$.
 In this unit physical quantities can be recovered by the relations
   $\bar{V}={V} c^2$, $\bar{E}={E}mc^2$, $M_{bh}=\hbar c r_s/mG$ and so on.

If there is no central BH (i. e., $r_s=0$),  a natural boundary condition is $d{\psi}/d{r}|_{{r}= 0}=0$,
 which gives the core-like density profile observed in dwarf galaxies \cite{Fuzzy,Matos:2003pe}.
On the other hand, if there is a central BH, the gravity of the BH changes the boundary condition of the field
at the center and gives a cuspy central density.
From the first equation of Eq. (\ref{SPE2}) one can see that as $r\rightarrow 0$, only $1/{r}$ dependent terms dominate and they give
a different boundary condition \cite{UrenaLopez:2002du}
\beq
  {\partial_{{r}}{\psi}}|_{{r}_0}=-{r}_s {{\psi}},
 \eeq
which means the bigger the black hole is,  the steeper the central field profile slope is.
(Here ${r}_0$ is a central point for the boundary condition where we can  use the Newtonian approximation while the gravity of the BH still influences the DM
scalar field.)
Though $\bar{r}_s$ is usually very small,
this small change of the boundary condition at the center could  result in a huge change in the overall DM density profile
due to the wave nature of the BEC/SFDM.
This phenomenon might be the key to understand the M-sigma relation.

We arbitrary choose  ${r}_0=100 {r}_s $ in this paper, and we focus on the
region where $ {r} \ge {r}_0$ so that we do not need to care about any relativistic effect or
details of BH physics.
This means that the effect of the BH on the relation is simply to give the boundary condition for the SFDM, which also implies that
any  super massive compact object at the galactic center  has a similar relation.

Since there is no known analytic solution of the SPE in Eq. (\ref{SPE2}),
we study approximate solutions and numerical solutions to see the effect of the black holes.
The approximate solutions of the SPE we find are
\beqa
\label{Vd}
{V}_d(r)&\simeq& {V}_0+\frac{{\psi}_0^2 {r}^2}{6},  \\
{\psi(r)}&\simeq& {\psi}_0 \left(1-{r}_s   {r}+ \frac{ ({V}_0-{E})}{3}  {r}^2\right),
\label{psi}
\eeqa
which can be checked by inserting them into the SPE. For example,
 $\nabla _{{r}} ^2 ({{V}_d -\frac{{r}_s}{{r}}}) \simeq {\psi_0}^2$ \cite{UrenaLopez:2002du}.

From Eq. (\ref{psi}) one can find that ${\psi}=\psi_0/2$ at
\beq
{r}_h=\frac{\sqrt{{r}_s^2 +2 w}-{r}_s}{2w},
\label{rh}
\eeq
where $w=(E-{V}_0)/3>0$. If the virial theorem holds $w=|V_0|/6$, and $r_h\simeq \sqrt{3/|V_0|}-3 r_s/|V_0|$.
${r}_h$ is   roughly the size of the bulge, which is a decreasing function of ${r}_s$.
For a  fixed  total mass $M\propto \int d^3 {r} |{\psi}|^2$, ${\psi}_0$ should increase
to compensate the reduction of the halo size.
This means that
the halo with a big BH has a narrow wave function and hence a compact DM density profile and a compact bulge.
(See Fig. \ref{figpsir})
Interestingly, this behavior is also consistent with observations of galaxies.
Graham et al. \cite{1538-3881-122-4-1707} have found that the
bulge light concentration  $C_{re}$ positively correlates with
$\sigma$ and $M_{bh}$, which means that a galaxy with a big BH has a compact bulge.
 The correlation is as strong as the M-sigma relation,
which implies they have a common physical origin. This fact supports our hypothesis.

Let us roughly derive the M-sigma relation using the approximate solutions.
The stellar rotation velocity is proportional to the square root of the potential depth.
From Eq. (\ref{psi}) we can define a size of the halo $r_f\simeq 1/\sqrt{w}=\sqrt{6/|V_0|}$
 using the condition $\psi(r_f)=0$.
 From the condition $V_d(r_f)=0=V_0+\psi_0^2/|V_0|$ one can see $\psi_0=|V_0|$.
  In Eq. (23) of Ref. \citealp{UrenaLopez:2002du} it was shown that $\psi_0\propto r_s^{1/2}$.
 As a result  $|V_0| \propto  r_s^{1/2}$, and hence
 $\sigma^4 \propto  |V_0|^2 \propto r_s$, which means
 $\beta=4$. However, this is a rough estimation based on the approximation.

\section{Numerical solutions}

To be more precise we performed a numerical calculation with Eq. (\ref{SPE2}) and the
 aforementioned boundary conditions using the shooting method~\cite{myhalo}. Another boundary condition is  $\psi\rightarrow  0$  as $r\rightarrow \infty$.
Owing to the virial theorem the star velocity dispersion can be given by $\sigma\simeq f v_{rot}({r}_h)$,
where $f$ is a model-dependent constant for which we take $1$ in this paper, and
$ v_{rot}({r}_h)$  is the stellar  rotation velocity at   ${r}_h$ defined by
\beq
 v_{rot}({r}_h)
\equiv ({V}_d({r}_h)-{V}_d({r}_0))^{1/2}c.
\eeq
In this equation we have assumed that the rotation velocity   of visible matter is determined by the depth of the potential well generated only by the DM. This approximation can be justified because the sphere of influence of the BH is usually small compared to the bulges
 and almost all visible matter in the bulges
remains far from the BH and hardly falls into the BH potential trap. (However, this approximation fails for $M_{bh}\simeq M$.)

\begin{figure*}
     \begin{minipage}[t]{0.3\linewidth}
      \centering\includegraphics[angle=0,width=5cm]{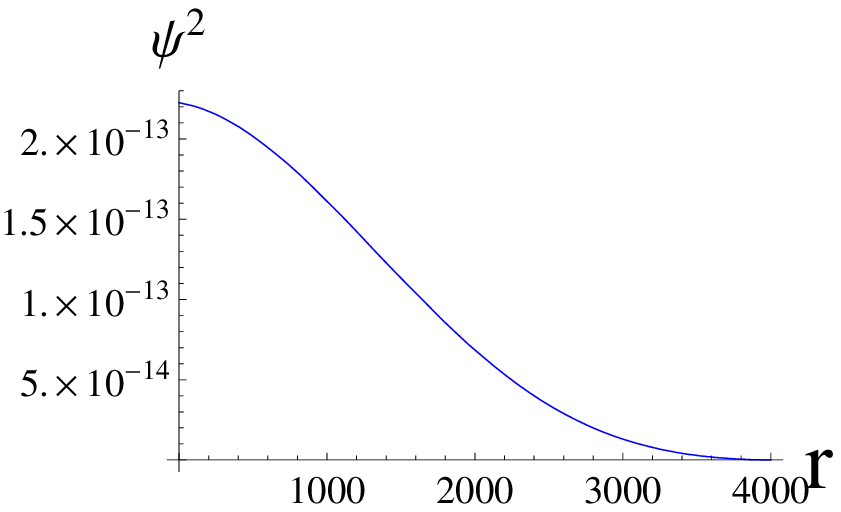}
     \hspace{0.1cm}
    \end{minipage}
 \begin{minipage}[t]{0.3\linewidth}
      \centering\includegraphics[angle=0,width=5cm]{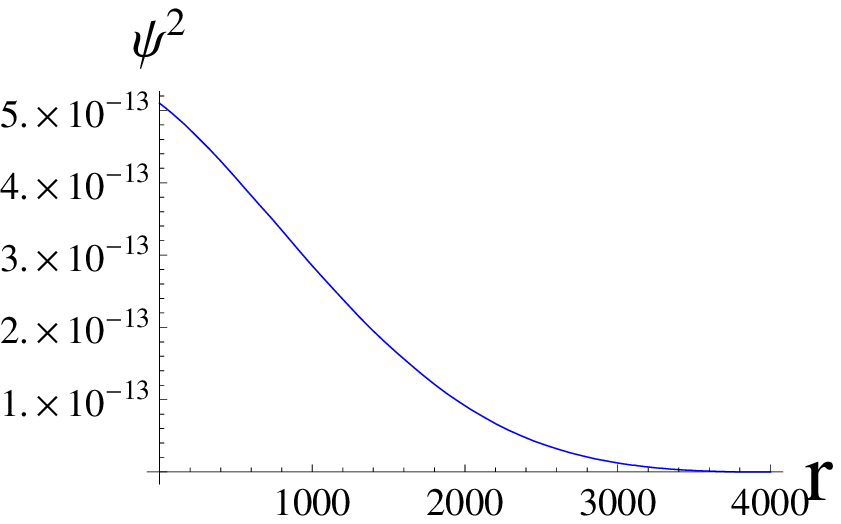}
     \hspace{0.1cm}
    \end{minipage}
 \begin{minipage}[t]{0.3\linewidth}
      \centering\includegraphics[angle=0,width=5cm]{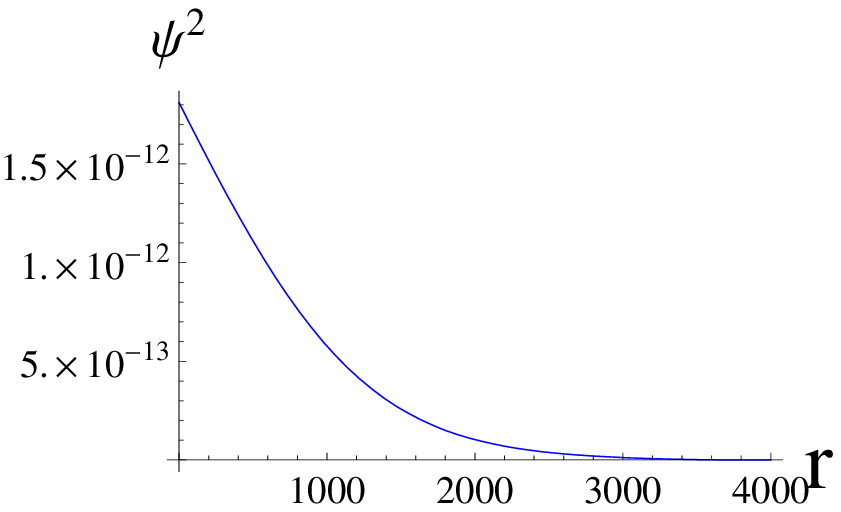}
     \hspace{0.1cm}
    \end{minipage}
 \caption{The dimensionless DM density profile versus $r$ depending on the BH mass.
 From the left $M_{bh}=9.3\times 10^6 M_\odot, 4.27\times 10^7 M_\odot$, and $1.1\times 10^8 M_\odot$ for $m=5\times 10^{-22} eV$, respectively.
For this $m$ the  unit length ($r=1$) is $0.0127pc$. The galaxy with a heavier BH has a more cuspy halo. }
 \label{figpsir}
\end{figure*}

\begin{figure*}
     \begin{minipage}[t]{0.3\linewidth}
      \centering\includegraphics[angle=0,width=5cm]{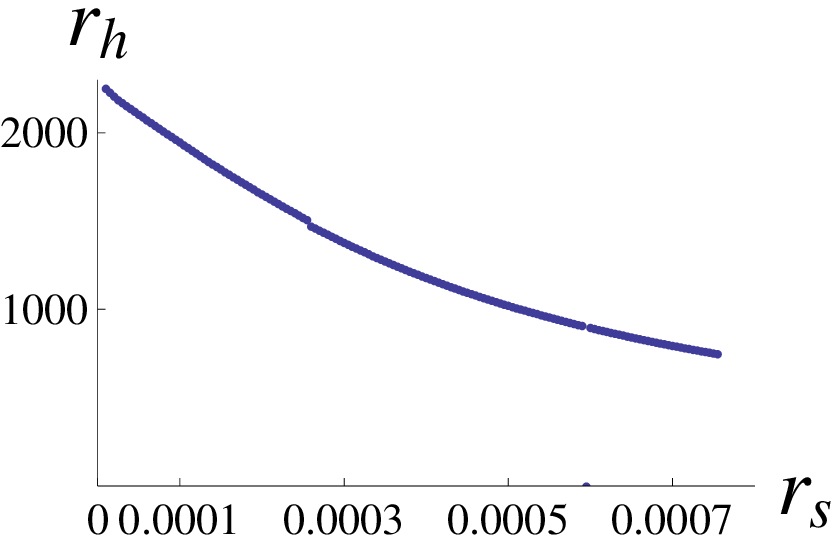}
     \hspace{0.1cm}
    \end{minipage}
 \begin{minipage}[t]{0.3\linewidth}
      \centering\includegraphics[angle=0,width=5cm]{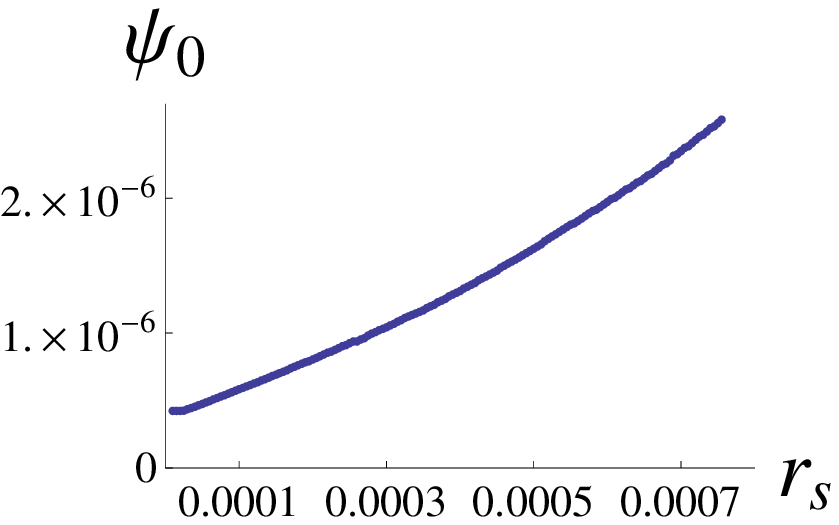}
     \hspace{0.1cm}
    \end{minipage}
 \caption{The effective radius of bulges (left) and the central field value (right) versus $r_s$.
 The heavier BH is, the more compact the bulges are.}
 \label{figrh}
\end{figure*}

 Figs. \ref{figpsir}-\ref{figrh} show the result of our numerical study with the parameters
 $V_d(r_0)\simeq  -4.3\times 10^{-7}$ and $E=-4.7\times 10^{-8}$.
The  total mass $M$  within $r=4000$ including the BH is  $O(10^8 M_\odot)$ for all 3 cases  with $m=5\times 10^{-22}eV$.
Fig. \ref{figpsir} shows the rescaled DM density profile as a function of $r$ for several different $M_{bh}$.
The graphs show  the effect of the BH on the scalar field.
The heavier BH is, the steeper the central slope of the DM density is.
  Fig. \ref{figrh} clearly reveals this tendency.
 The effective radius $r_h$ of the bulges is inversely correlated with $r_s$
 as predicted in Eq. (\ref{rh}), and the central field value increases as $r_s$.
  We assumed $\psi_0\simeq\psi(r_0)$ for the interpretation of the  numerical work.
 Therefore, the heavier BH is, the more compact the bulge is, and
  more compact bulges means a larger $\sigma$, and hence the M-sigma relation.
The numerical  results support the semi-analytical arguments above.

\begin{figure}[htbp]
\includegraphics[width=0.4\textwidth]{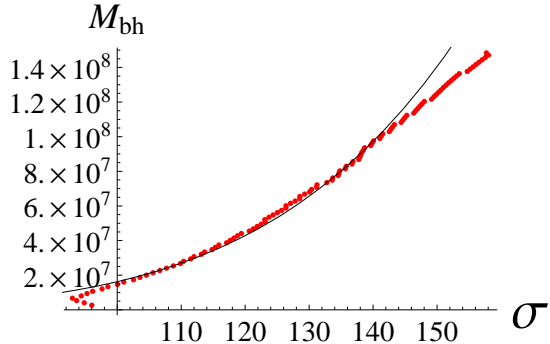}
\caption{(Color online)
The black hole mass (in $M_\odot$) versus  the velocity dispersion (in $km/s$) of  model galaxies  (red dots) for  $m = 5\times 10^{-22}eV$.
The best fitting curve (black line) gives $\alpha=-3.4$ and $\beta=5.3$ which are consistent with the observational
data.
Each point represents a galaxy.
}
\label{figmsigma}
\end{figure}

The numerical solutions of the dimensionless SPE  are independent of $m$.
To compare the results with observations we first need to fix $m$.
The  particle mass   $m\simeq 10^{-22} eV$ is required to solve
the cusp problem and to suppress the small-scale power \cite{Fuzzy},
and in Ref. \citealp{Lee:2008jp}  it was shown that the SFDM with $m\simeq 5\times 10^{-22} eV$ can explain the minimum mass and the size of galaxies.
Let us use this $m$ for the fitting below.

Fig. \ref{figmsigma} shows our $M_{bh}-\sigma$ relation obtained from the numerical calculation.
We  check the relation with the  parameterization
\beq
log_{10} M_{bh}=\alpha + \beta log_{10} {\sigma},
\eeq
where $M_{bh}$ is in $M_\odot$ unit and $\sigma$ is in $kms^{-1}$.
There is still a  controversy about the value of the exponent $\beta$.
For example,  Gebhardt et al. \cite{1538-4357-539-1-L13}
suggested $\beta=3.75 \pm 0.3$
 while Ferrarese and Merritt \cite{Ferrarese:2000se}  had reported  $(\alpha,\beta)=(-2.9\pm 1.3,4.8\pm 0.54)$.
For  $m=5\times 10^{-22}eV$  and  $\beta=4.8$, we obtain the best fitting value $\alpha=-2.3$,
which is comparable to the observed value.
On the other hand,  our best fitting value for  free $(\alpha,\beta)$ is $(\alpha,\beta)=(-3.4,5.3)$, which is also quite  similar to the recently inferred value  $(\alpha,\beta)=(-3.4\pm 0.86,5.12\pm 0.36)$ ~\cite{2011Natur.480..215M}.
This is a remarkable consistency despite  a few free parameters and the simple arguments we have used so far.

Since $\alpha$ is very sensitive to $m$ while $\beta$ is not,  we can
 constrain $m$ from the fitting. For $m=10^{-22} eV$ our fitting gives $\alpha=-2.7$ while for $m=10^{-21}eV$ $\alpha=-3.7$.
 Therefore, it is  interesting that the DM particle mass $m\simeq 5\times 10^{-22} eV$ required to reproduce the correct $\alpha$
 is just that for explaining the minimum size and the mass of galaxies \cite{Lee:2008jp,Lee:2008ux}.
 The BEC/SFDM can explain not only the exponent but also the normalization of the M-sigma relation.

\section{Discussion}

The fitting in the previous section was done  only for the $\sigma$ range where the simple power-law holds.
For  $M_{bh}\ll M$ the effect of the BH on  $M_{bh}$  is negligible and
we expect a large scatter of the M-sigma relation as often found in the  M-sigma relation graphs \cite{2011Natur.480..215M}.
On the other hand, for $M_{bh}\sim M$ the gravitational force exerted by  the BH
is much larger than
that by the DM and we expect $\sigma$ to follow not the M-sigma relation but
 a  position dependent Kepler-law like velocity dispersion $\sigma\simeq \sqrt{G M_{bh}/r}$.
 Interestingly, the recent observation of the over-massive black hole in NGC 1277 \cite{2012Natur.491..729V} showed
 this behavior of $\sigma$.
In this galaxy the mass of
the central black hole ($M_{bh}=1.7\times 10^{10} M_\odot$) is $59\%$ of its bulge mass, which is quite different from the value
 $M_{bh}\simeq 2\times 10^9 M_\odot$ estimated from the M-sigma relation.
  In our model  the simple power-law M-sigma relation is no longer valid for
a relatively large or small $M_{bh}$ compared to $M$ as expected theoretically. This is different from
the prediction of the feedback mechanisms proposed so far.
 The numerical result in Fig. \ref{figmsigma} also shows this tendency.
 To study the effect of very heavy  SMBHs we need a SFDM halo profile for a  big galaxy, which is unclear yet.

In conclusion we have shown  that BEC/SFDM model with $m=5\times 10^{-22}eV$ can explain the  M-sigma relation
between the SMBH mass and the velocity dispersions in  galaxies as well as other cosmological constraints.
This resolution does not need any feedback mechanism, and the M-sigma relation is robust against mergers, and hence universal.
Since the M-sigma relation is established by the DM wave dynamics, the tight relation is automatically and
rapidly readjusted after the mergers.
This model also  seems to  explain the relation between  the bulge size and the BH mass.
Therefore, the BEC/SFDM model seems to give a new hint to the BH-galaxy coevolution.
More detailed analytical and numerical studies including visible matter are desirable from this perspective.
\\

\section*{ ACKNOWLEDGMENTS }

\vskip 5.4mm
%
%

\begin{thebibliography}{10}

\bibitem{Merritt:1999ry}
D. Merritt, ASP Conf. Ser. {\bf 197},  221  (2000).

\bibitem{Ferrarese:2000se}
L. Ferrarese and D. Merritt, Astrophys. J. {\bf 539},  L9  (2000).

\bibitem{Silk:1997xw}
J. Silk and M.~J. Rees, Astron. Astrophys. {\bf 331},  L1  (1998).

\bibitem{King:2003ix}
A. King, Astrophys. J. {\bf 596},  L27  (2003).

\bibitem{1983PhLB..122..221B}
M.~R. {Baldeschi}, G.~B. {Gelmini}, and R. {Ruffini}, Physics Letters B {\bf
  122},  221  (1983).

\bibitem{1989PhRvA..39.4207M}
M. {Membrado}, A.~F. {Pacheco}, and J. {Sa{\~n}udo}, \pra {\bf 39},  4207
  (1989).

\bibitem{sin1}
S.-J. Sin, Phys. Rev. {\bf D50},  3650  (1994).

\bibitem{myhalo}
J.-W. Lee and I.-G. Koh, Phys. Rev. {\bf D53},  2236  (1996).

\bibitem{Schunck:1998nq}
F.~E. Schunck, astro-ph/9802258  (1998).

\bibitem{Matos:1998vk}
T. Matos and F.~S. Guzman, Class. Quant. Grav. {\bf 17},  L9  (2000).

\bibitem{PhysRevLett.84.3037}
U. Nucamendi, M. Salgado, and D. Sudarsky, Phys. Rev. Lett. {\bf 84},  3037
  (2000).

\bibitem{PhysRevD.64.123528}
A. Arbey, J. Lesgourgues, and P. Salati, Phys. Rev. D {\bf 64},  123528
  (2001).

\bibitem{repulsive}
J. Goodman, New Astronomy Reviews {\bf 5},  103  (2000).

\bibitem{Fuzzy}
W. Hu, R. Barkana, and A. Gruzinov, Phys. Rev. Lett. {\bf 85},  1158  (2000).

\bibitem{corePeebles}
P. Peebles, \apj {\bf 534},  L127  (2000).

\bibitem{Nontopological}
E.~W. Mielke and F.~E. Schunck, Phys. Rev. D {\bf 66},  023503  (2002).

\bibitem{PhysRevD.62.103517}
V. Sahni and L. Wang, Phys. Rev. D {\bf 62},  103517  (2000).

\bibitem{Alcubierre:2001ea}
M. Alcubierre {\it et~al.}, Class. Quant. Grav. {\bf 19},  5017  (2002).

\bibitem{Fuchs:2004xe}
B. Fuchs and E.~W. Mielke, Mon. Not. Roy. Astron. Soc. {\bf 350},  707  (2004).

\bibitem{Matos:2001ps}
T. Matos, F.~S. Guzman, L.~A. Urena-Lopez, and D. Nunez, astro-ph/0102419  .

\bibitem{0264-9381-18-17-101}
M.~P. Silverman and R.~L. Mallett, Classical and Quantum Gravity {\bf 18},
  L103  (2001).

\bibitem{PhysRevD.63.125016}
U. Nucamendi, M. Salgado, and D. Sudarsky, Phys. Rev. D {\bf 63},  125016
  (2001).

\bibitem{Julien}
A.~A. Julien~Lesgourgues and P. Salati, New Astronomy Reviews {\bf 46},  791
  (2002).

\bibitem{moffat-2006}
J.~W. Moffat, astro-ph/0602607  (2006).

\bibitem{2011PhRvD..84d3531C}
P.-H. {Chavanis}, \prd {\bf 84},  043531  (2011).

\bibitem{2009JKPS...54.2622L}
J.-W. {Lee}, Journal of Korean Physical Society {\bf 54},  2622  (2009).

\bibitem{2014ASSP...38..107S}
A. {Su{\'a}rez}, V.~H. {Robles}, and T. {Matos}, Astrophysics and Space Science
  Proceedings {\bf 38},  107  (2014).

\bibitem{2014MPLA...2930002R}
T. {Rindler-Daller} and P.~R. {Shapiro}, Modern Physics Letters A {\bf 29},
  30002  (2014).

\bibitem{2014PhRvD..89h4040H}
T. {Harko}, \prd {\bf 89},  084040  (2014).

\bibitem{2014IJMPA..2950074H}
K. {Huang}, C. {Xiong}, and X. {Zhao}, International Journal of Modern Physics
  A {\bf 29},  50074  (2014).

\bibitem{2010arXiv1004.4016B}
H.~L. {Bray}, arXiv1004.4016  (2010).

\bibitem{PhysRevD.63.063506}
T. Matos and L. Arturo Ure\~na L\'opez, Phys. Rev. D {\bf 63},  063506  (2001).

\bibitem{Salucci:2002nc}
P. Salucci, F. Walter, and A. Borriello, Astron. Astrophys. {\bf 409},  53
  (2003).

\bibitem{navarro-1996-462}
J.~F. Navarro, C.~S. Frenk, and S.~D.~M. White, \apj {\bf 462},  563  (1996).

\bibitem{deblok-2002}
W.~J.~G. {de Blok}, A. Bosma, and S.~S. McGaugh, astro-ph/0212102  (2002).

\bibitem{crisis}
A. Tasitsiomi, International Journal of Modern Physics D {\bf 12},  1157
  (2003).

\bibitem{0264-9381-17-1-102}
F.~S. Guzman and T. Matos, Class. Quant. Grav. {\bf 17},  L9  (2000).

\bibitem{Mbelek:2004ff}
J.~P. Mbelek, Astron. Astrophys. {\bf 424},  761  (2004).

\bibitem{PhysRevD.69.127502}
T.~H. Lee and B.~J. Lee, Phys. Rev. D {\bf 69},  127502  (2004).

\bibitem{2014NatPh..10..496S}
H.-Y. {Schive}, T. {Chiueh}, and T. {Broadhurst}, Nature Physics {\bf 10},  496
   (2014).

\bibitem{review}
P. Jetzer, Phys. Rep. {\bf 220},  163  (1992).

\bibitem{Schive:2014hza}
H.-Y. Schive {\it et~al.}, Phys. Rev. Lett. {\bf 113},  261302  (2014).

\bibitem{Matos:2003pe}
T. Matos and D. Nunez, astro-ph/0303455  (2003).

\bibitem{UrenaLopez:2002du}
L.~A. Urena-Lopez and A.~R. Liddle, Phys. Rev. {\bf D66},  083005  (2002).

\bibitem{1538-3881-122-4-1707}
A.~W. Graham, I. Trujillo, and N. Caon, The Astronomical Journal {\bf 122},
  1707  (2001).

\bibitem{Lee:2008jp}
J.-W. Lee and S. Lim, JCAP {\bf 1001},  007  (2010).

\bibitem{1538-4357-539-1-L13}
K. Gebhardt {\it et~al.}, The Astrophysical Journal Letters {\bf 539},  L13
  (2000).

\bibitem{2011Natur.480..215M}
N.~J. {McConnell} {\it et~al.}, \nat {\bf 480},  215  (2011).

\bibitem{Lee:2008ux}
J.-W. Lee, Phys. Lett. {\bf B681},  118  (2009).

\bibitem{2012Natur.491..729V}
R.~C.~E. {van den Bosch} {\it et~al.}, \nat {\bf 491},  729  (2012).

\end{thebibliography}

\end{document}